\documentclass[aps,prx,twocolumn,footinbib,showpacs,showkeys]{revtex4-1}
\usepackage{graphicx}
\usepackage{indentfirst}
\usepackage{physics}
\usepackage{braket}
\usepackage{float}
\usepackage{amsmath}
\usepackage{epstopdf}
\usepackage{CJK}
\usepackage{esint}
\usepackage{color}
\usepackage[T1]{fontenc}
\usepackage{subfigure}
\usepackage{amsfonts}
\usepackage{footmisc}
\usepackage{scrextend}
\usepackage{multirow}
\usepackage[hyperfootnotes=false]{hyperref}

\usepackage[english]{babel}
\usepackage{url}
\usepackage{bm}
\usepackage{hyperref}
\definecolor{darkblue}{rgb}{0,0,0.5}
\hypersetup{
colorlinks=true,
linkcolor=black,
filecolor=blue,
citecolor=darkblue,  
urlcolor=black,
}

\newcommand{\calE}{{\cal E}}

\newcommand{\1}{^{(1)}}

\def\be{\begin{equation}}
\def\ee{\end{equation}}
\def\ba{\begin{eqnarray}}
\def\ea{\end{eqnarray}}
\usepackage{bm}

\begin{document}

\title{{Physical-Layer} Supervised Learning Assisted by an Entangled Sensor Network}

\author{Quntao Zhuang$^{1,2,4}$}
\email{zhuangquntao@gmail.com}
\author{Zheshen Zhang$^{3,4}$}
\affiliation{
$^1$Department of Electrical and Computer Engineering, University of Arizona, Tucson, Arizona 85721, USA
\\
$^2$Department of Physics, University of California, Berkeley, California 94720, USA
\\
$^3$Department of Materials Science and Engineering, University of Arizona, Tucson, Arizona 85721, USA
\\
$^4$James C. Wyant College of Optical Sciences, University of Arizona, Tucson, AZ 85721, USA
}
\date{\today}

\begin{abstract} 
Many existing quantum supervised learning (SL) schemes consider data given {\em a priori} in a classical description. With only noisy intermediate-scale quantum (NISQ) devices available in the near future, their quantum speedup awaits the development of quantum random access memories (qRAMs) and fault-tolerant quantum computing. There, however, also exist a multitude of SL tasks whose data are acquired by sensors, e.g., pattern classification based on data produced by imaging sensors. Solving such SL tasks naturally requires an integrated approach harnessing tools from both quantum sensing and quantum computing. We introduce supervised learning assisted by an entangled sensor network (SLAEN) as a means to carry out SL tasks at the physical layer. The entanglement shared by the sensors in SLAEN boosts the performance of extracting global features of the object under investigation. We leverage SLAEN to construct an entanglement-assisted support-vector machine for data classification and entanglement-assisted principal component analyzer for data compression. In both schemes, variational circuits are employed to seek the optimum entangled probe states and measurement settings to maximize the entanglement-enabled {enhancement}. We observe that SLAEN enjoys an appreciable entanglement-enabled performance gain, even in the presence of loss, over conventional strategies in which classical data are acquired by separable sensors and subsequently processed by classical SL algorithms. SLAEN is realizable with available technology, opening a viable route toward building NISQ devices that offer unmatched performance beyond what the optimum classical device is able to afford.

\end{abstract} 


\maketitle
\section{Introduction}
While error-corrected scalable quantum computation is not yet available, recent advances towards small-scale quantum computers~\cite{IBMQ,castelvecchi2017quantum} have spurred interests in seeking noisy intermediate-scale quantum (NISQ)~\cite{Preskill2018quantumcomputingin} devices that offer unmatched performance beyond the capability of classical devices. The theoretical proof of a quantum advantage over classical computation~\cite{bravyi2018quantum,bouland2018complexity} further inspires the search for quantum-assisted schemes tailored for practical tasks. In this regard, quantum variational schemes, in conjunction with classical optimization, are widely applicable to tasks including quantum state preparation~\cite{wecker2015progress}, variational eigensolvers~\cite{peruzzo2014variational,kandala2017hardware}, state diagonalization~\cite{larose2018variational}, and machine learning~\cite{havlicek2018supervised,schuld2018quantum,biamonte2017quantum,dunjko2018machine,killoran2018continuous,rebentrost2018,lloyd2018quantum,Steinbrecher_2018,schuld2018,schuld2015introduction,PerdomoOrtiz2018}. 

Various quantum machine-learning algorithms rest upon input data given {\em a priori} in a classical description, e.g., data from statistics of online users' activities, social networks, and financial transactions. To carry out quantum machine learning, classical data are quantum encoded, stored in quantum random access memories (qRAMs), and processed by quantum circuits. These quantum machine-learning algorithms aim to efficiently solve certain tasks that are intractable by the optimum classical algorithm, but achieving such a quantum advantage is held back by outstanding challenges associates with, e.g., preserving quantum coherence, efficient construction of qRAMs~\cite{arunachalam2015robustness}, and a lack of understanding for what noisy quantum circuits would be able to offer. There are, however, also many situations in which the input data are acquired through quantum measurements. Such a scenario creates opportunities to harness tools grown out of quantum sensing to assist machine-learning tasks and achieve a performance gain over conventional approaches based on classical machine-learning algorithms supplied with massive sensor-produced data.

In this paper, we introduce supervised learning assisted by an entangled sensor network (SLAEN), a paradigm to assist supervised learning (SL) tasks at the physical layer through the sensing process that acquires data. As Sec.~\ref{data_acquisition} elucidates, while the current form of SLAEN is unable to assist SL tasks based on classical data given {\em a priori}, it can indeed yield an advantage over classical SL schemes that rely on data acquired by separable sensors. The SLAEN architecture consists of quantum circuits and an entangled sensor network. Utilizing a variational approach, the quantum circuits optimize the multipartite entangled probe state shared in the sensor network and seek the optimum measurement setting to capture global features of interest. The SLAEN architecture can be employed in two major tasks, a support-vector machine (SVM) for data classification and a principal component analyzer (PCA) for data compression. Because SLAEN carries out SL tasks directly at the physical layer through quantum measurements, its entanglement-enabled enhancement over classical SL schemes without entanglement can be quantified using tools from quantum measurement theory. Specifically, in classification tasks undertaken by $M$ entangled sensor nodes, the error probability of SLAEN $\propto P^M$ can be exponentially lower than the error probability $P$ of the optimum classical strategy, {owing to the entanglement-enabled Heisenberg scaling in measurement sensitivity~\cite{giovannetti2006,giovannetti2011advances,pirandola2018}.} {For general classification tasks on random data, we give numerical evidences for a substantial performance improvement.} SLAEN can be readily implemented by off-the-shelf components such as single-mode squeezers, linear optical circuits, and homodyne measurements to carry out SL tasks pertaining to non-demolition detection~\cite{eckert2008quantum}, thermometry~\cite{mehboudi2015thermometry}, radial-frequency (RF) sensing~\cite{fan2018superconducting}, biomolecule sensing~\cite{fernandez2016last}, and phase estimation~\cite{escher2011general,demkowicz2013fundamental}.

\section{Problem description}
\label{data_acquisition}
Many existing SL tasks rely on classical data available {\em a priori}. Examples of such tasks include classification on data collected from surveys, recommendation systems based on online user statistics, and market predictions made from historical financial data. To achieve a quantum speedup for such problems, a typical approach is loading classical data into qRAMs followed by quantum processing~\cite{biamonte2017quantum, dunjko2018machine, rebentrost2018, lloyd2018quantum, schuld2018, schuld2015introduction, PerdomoOrtiz2018}. With only NISQ devices available in the near future, however, a quantum speedup for such SL tasks is held back by the immense technical challenges in building qRAMs and fault-tolerant quantum computing.

\begin{figure}
\centering
\includegraphics[width=0.48\textwidth]{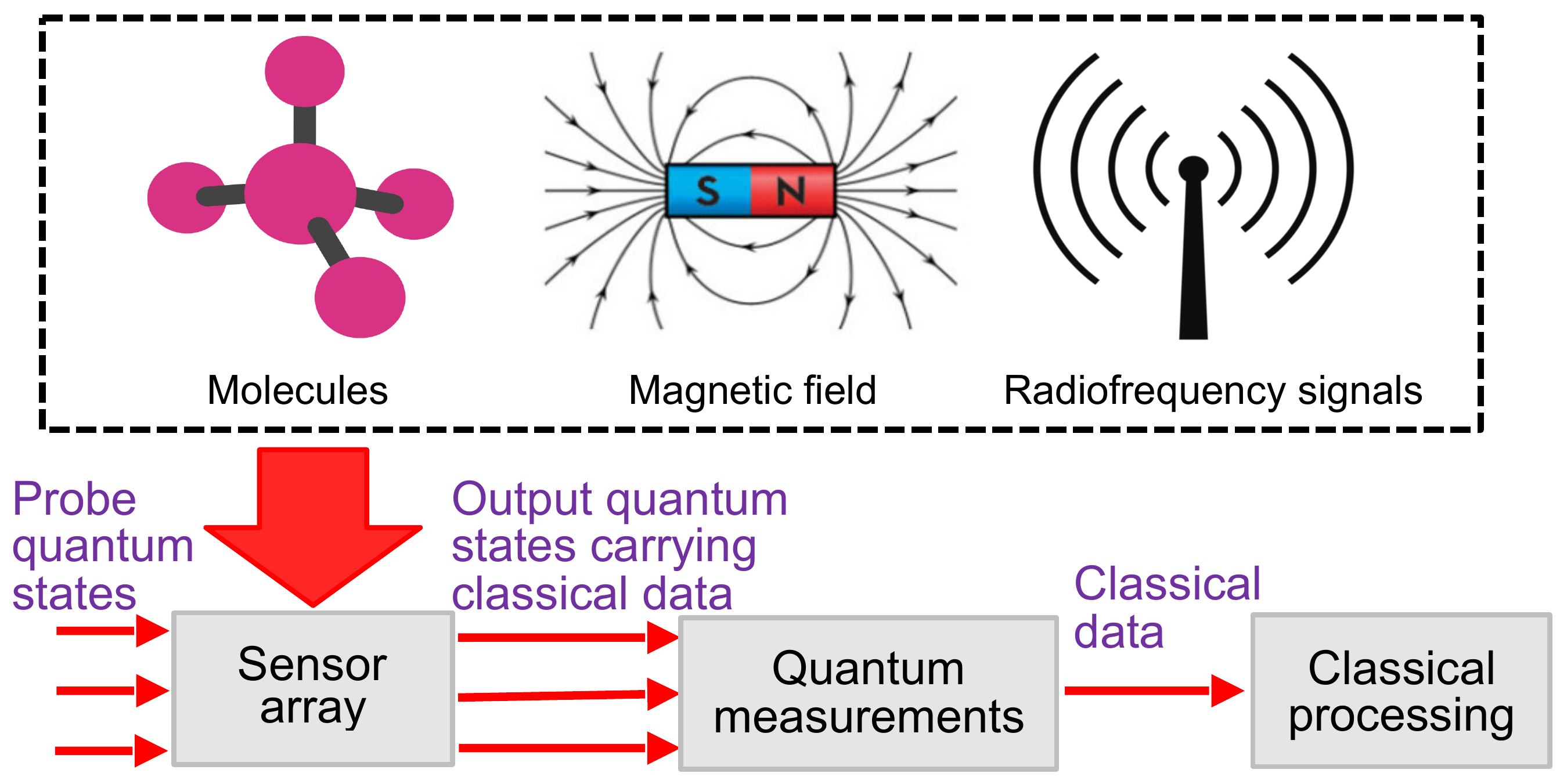}
\caption{
  Schematic of supervised learning based on data generated from sensors. The output quantum states carry classical data about the object under investigation.
  \label{fig:data_acquisition}
}
\end{figure}

There are, however, an abundance of SL tasks in which classical data are not available until they are acquired by sensors. For example, biosensors probe an unknown specimen to generate data that are post-processed for molecular identification~\cite{fernandez2016last}; an array of charge-coupled-device sensors captures an image of an object to perform classification; fiber optic gyroscopes produce rotational data for navigation; radiofrequency sensors measure instantaneous frequencies for signal analysis~\cite{maram2019}; and magnetometers sense magnetic fields to investigate the brain~\cite{hamalainen1994}. From a quantum-information perspective, a general schematic for a SL task involving sensing to acquire data is illustrated in Fig.~\ref{fig:data_acquisition}. Probe quantum states are first initialized and then altered by an array of sensors that probe an object of interest. The classical data about the object are carried by the output quantum states of the sensor array. Next, quantum measurements are taken to retrieve the classical data, and, finally, classical processing on the classical data completes the SL task. Two important questions arise in such a configuration. First, can entanglement shared by the probe quantum states assist a SL task and improve its performance? Second, can a SL task be carried out at a physical layer via choosing proper probe quantum states and measurement settings so that the resource overhead on classical processing is reduced? SLAEN answers yes to both questions.

In the following Sections, we will show that while SLAEN, in its present form, does not benefit SL tasks based on classical data given {\em a priori}, it is indeed a useful technique in the NISQ era to assist SL tasks involving sensing through optimizing the entangled probe states and joint measurements. Before going into details of SLAEN, here we point out an important departure of SLAEN's utility from previous works. In various quantum machine learning works mentioned above, either the data is classically given and one utilizes quantum enhancement to perform the learning; or the problem is entirely quantum in the sense that the initial data are unknown quantum states. In our scenario, the data are unknown classical information, and a sensor transforms the classical unknown data into quantum states for further processing. Such a scenario, albeit ubiquitous for sensing applications, has not been explored in quantum machine learning.

\section{The architecture}

\begin{figure}
\centering
\includegraphics[width=0.48\textwidth]{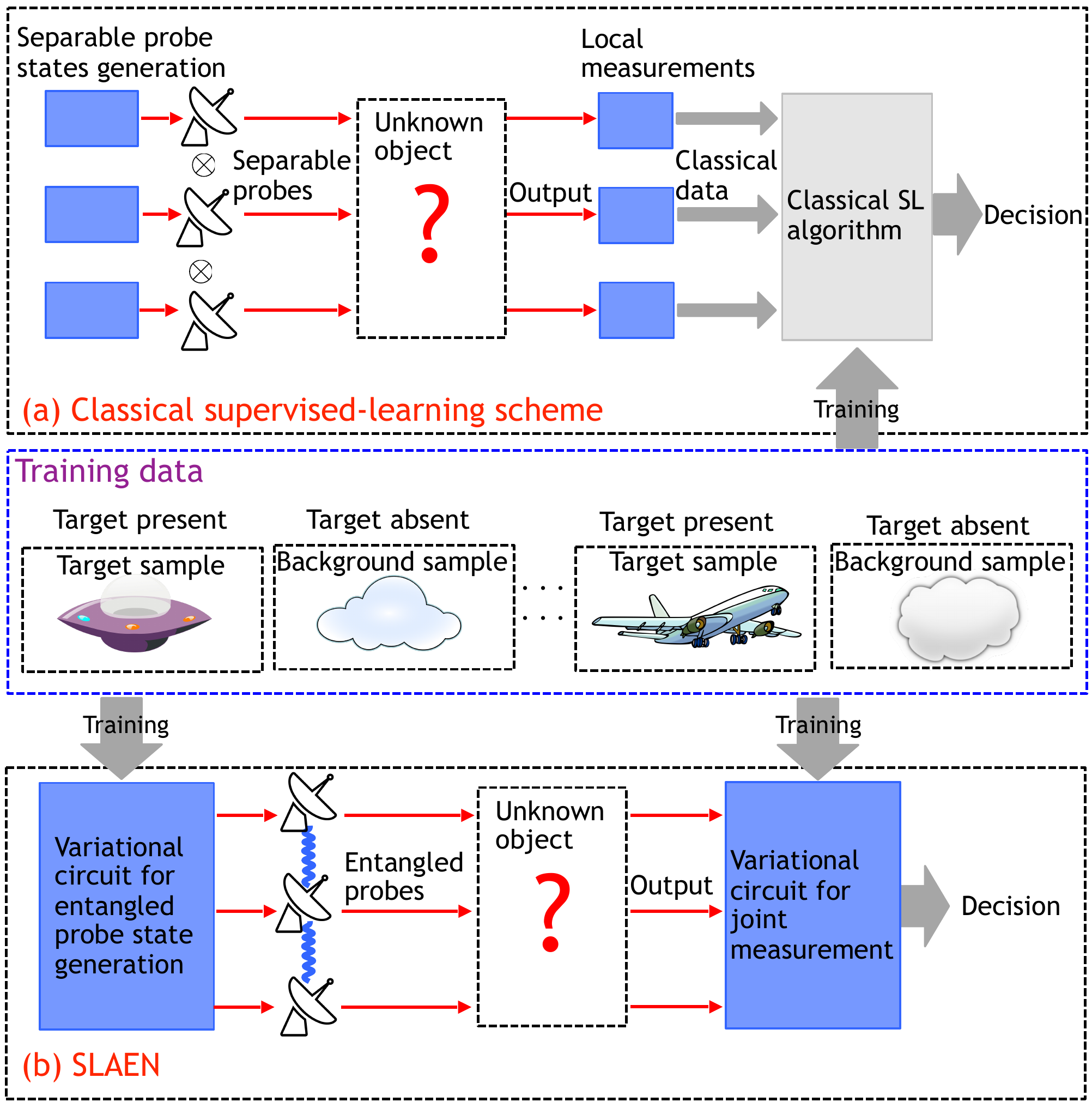}
\caption{\label{Fig0}
A classical SL scheme versus SLAEN in a data classification task for target detection.
(a) In a classical SL scheme without entanglement, the separable sensor network is connected with a classical SL algorithm. Separable probe states and local measurements are employed to investigate an unknown object. The classical data acquired by measurements are fed into a classical SL algorithm to make a decision. 
(b) In SLAEN, the sensors, represented by the probe icons, share a multipartite entangled probe state. Each sensor sends its share of the entangled probe state to investigate an unknown object. The sensing process is modeled as parameter estimation of quantum channels. Each sensor views the object from a different perspective, and their shared entangled probe state boosts the performance of estimating global parameters of the object. Prior to interrogating an unknown object, existing data are used to train the first variational circuit (left blue box) to generate the optimum entangled probe state and the second variational circuit (right blue box) to perform the optimum measurement. 
}
\end{figure}

From a quantum-sensing perspective~\cite{giovannetti2006,giovannetti2011advances,pirandola2018}, data acquisition is modeled as parameter estimation of a channel, i.e., a completely-positive and trace-preserving (CPTP) linear map~\cite{caruso2014quantum}, using a probe state followed by quantum measurements. For instance, in the target-detection example illustrated in Fig.~\ref{Fig0}, interrogating the presence or absence of the target is through using a probe state to estimate the transmissivity of a lossy and noisy bosonic channel~\cite{Tan2008}. Such a generalized data-acquisition picture is applicable to wide range of sensing applications such as fluorescence imaging~\cite{denk1990two,eshun2018investigations}, optical reading~\cite{pirandola2011quantum}, phase sensing~\cite{escher2011general,demkowicz2013fundamental,zhuang2018distributed}, magnetic measurements~\cite{maze2008nanoscale}, and fiber optic gyroscopes~\cite{bergh1981all}.

{\em Classical SL schemes without entanglement.---}The amount of available information, carried on the probe state, about the object under investigation is ultimately limited by the quantum Cram\'{e}r-Rao bound~\cite{Helstrom_1976,Holevo_1982,Yuen_1973} arising from measurement uncertainties. 
Thus, the probe state set a fundamental performance limit for its associated SL task. As illustrated in Fig.~\ref{Fig0}(a), a conventional classical SL scheme, like what is {\em standard} with imaging sensors, uses separable probe states and make separable local measurements on the output states to obtain classical data that describe the object under investigation. The classical data are then fed into a classical SL algorithm, trained by known data, to extract features of the interrogated object. Since most SL tasks only aim to investigate global features of an object, obtaining a full description through separable local measurements and passing all the classical data to a classical SL algorithm is not only redundant but also places a substantial resource overhead.

{\em SLAEN.---}
To overcome this overhead, SLAEN carries out SL tasks at a physical layer through the data-acquiring sensing process, without resorting to a classical SL algorithm that copes with massive classical data. At the core of the SLAEN architecture is a network of entangled sensors optimized by variational circuits, as illustrated by a target-detection example in Fig.~\ref{Fig0}(b). SLAEN's {performance enhancement} over the classical SL scheme described in Fig.~\ref{Fig0}(a) is two-fold: (i) the entanglement shared by different sensors reduces errors and boosts the sensitivity of extracting global information of the object under investigation, thereby leading to an {entanglement enhancement} in SL tasks, such as SVM and PCA, over the compared classical SL schemes; and (ii) direct extraction of global features by joint measurements optimized by the variational circuit obviates the overhead arising from obtaining redundant local information about the object;  SLAEN's capability of carrying out entanglement-assisted SL tasks at a physical layer makes it significantly different from existing quantum SL schemes that take classical data as their inputs.

Prior studies~\cite{zhuang2018distributed,ge2017distributed,proctor2017multi} on entangled sensor networks show that for the problem of estimating a global parameter of the network, e.g., a weighted sum of phase shifts measured by different sensors, shared entanglement leads to a measurement-sensitivity scaling advantage, i.e., approaching the Heisenberg limit~\cite{Giovannetti_2001,giovannetti2004,giovannetti2006,giovannetti2011advances,escher2011,zhuang2017entanglement} with respect to the number of sensors, over the optimum separable sensor network. In particular, Ref.~\cite{zhuang2018distributed} presents entangled sensor networks based on continuous variables (CVs) and proved that the root-mean-square (rms) error for estimating a weighted sum of displacements at different sensor nodes is a factor of $\sim 1/\sqrt{M}$ smaller than a product-state sensor network, where $M$ is the number of sensor nodes. In addition, Appendix~\ref{sec_DS} shows that the measurement-sensitivity advantage in parameter estimation translates to an {\em exponential} error-probability advantage in channel discrimination. 

Entangled sensor networks' advantage in channel discrimination makes them well suited for SL tasks. To fully unleash the power of the entangled sensor network for different SL tasks, one needs to tailor the entangled probe state shared between the sensors and optimize the measurement setting. For a complex SL task, however, deriving the optimum entangled probe state and measurement configuration becomes a formidable problem. Take the target-detection problem depicted in Fig.~\ref{Fig0} as an example. The target object is embedded in a highly dynamic environment so our goal is to optimize the entangled probe state shared by the sensor nodes and design a measurement strategy to minimize the error probability of interrogating the presence or absence of the target. In the SLAEN architecture, an optical circuit, trained by a variational approach, generates the entangled probe state. Likewise, a second variational optical circuit processes the output entangled probe state to perform a {\em joint} measurement that captures the object feature of interest. In doing so, SL tasks are carried out during the data-acquiring sensing process at a physical layer, thereby eliminating the need for precise knowledge about local features. The training data and the variational approach ensure that the {entanglement enhancement} over the compared classical SL schemes is reaped for a variety of SL tasks. Moreover, since CV entangled sensor networks enjoy loss tolerance, deterministic entanglement preparation, and efficient broadband measurements~\cite{zhuang2018distributed}, the {performance enhancement} of SLAEN over the classical SL schemes is achievable with available technology.

\section{Entanglement-assisted physical-layer support-vector machine}
\label{sec_classical_SVM}
\begin{figure}
\centering
\includegraphics[width=0.5\textwidth]{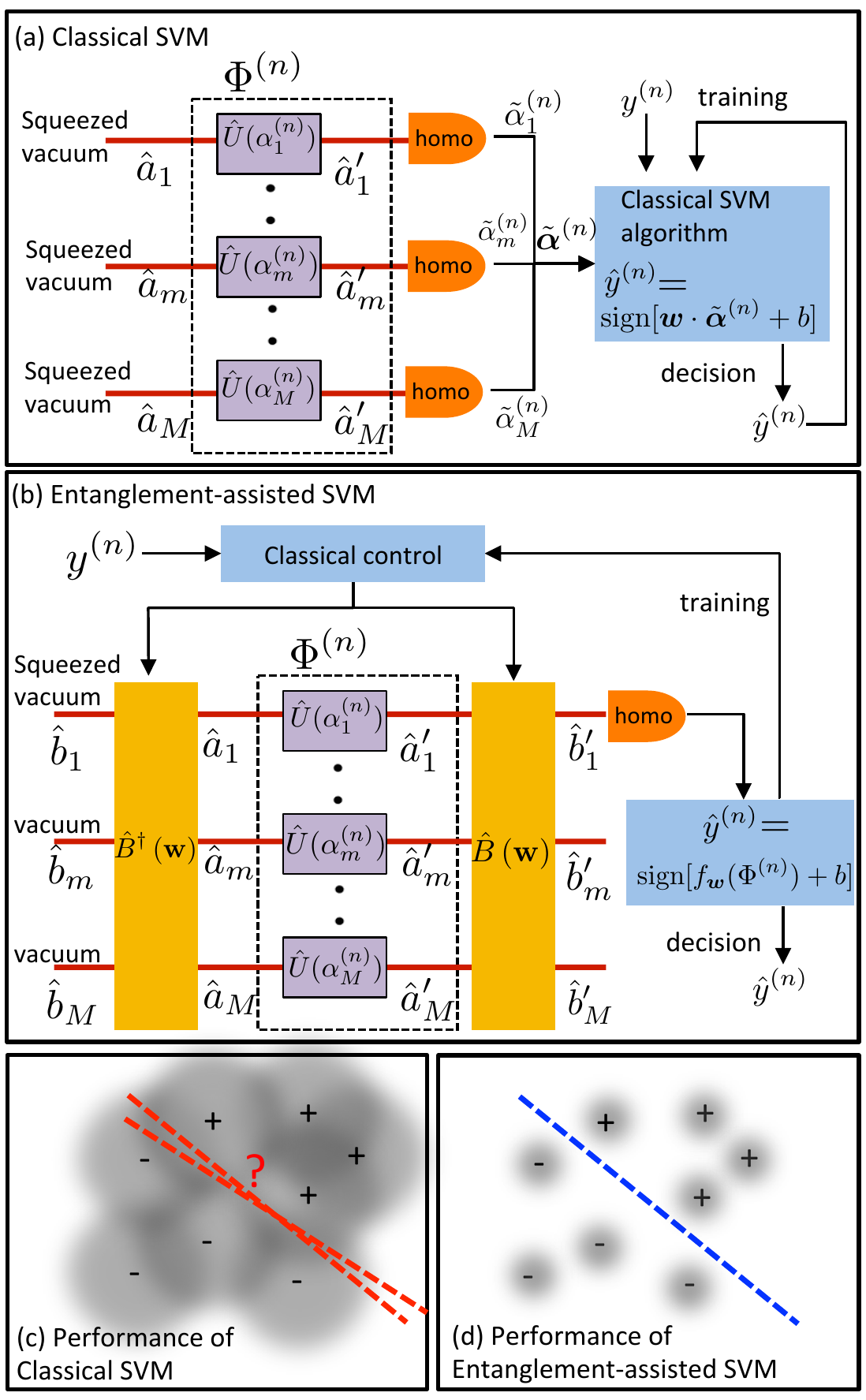}
\caption{Schematic of the classification task of an unknown quantum channel $\Phi^{(n)}$ comprised of $M$ displacement operations paremetrized by an unknown vector $\bm \alpha^{(n)} = \{\alpha^{(n)}_1, \cdots, \alpha^{(n)}_M\}$. ``homo'' denotes a homodyne measurement. (a) schematic of a classical SVM with separable probe states, which obtain full noisy classical description $\tilde{\bm \alpha}^{(n)}$ about the probed object and then feed it into a classical SVM algorithm. (b) schematic of the entanglement-assisted SVM. Two variational circuits $\hat{B}^\dag({\bm w})$ and $\hat{B}({\bm w})$ are trained to generate the optimum entangled probe state and to preprocess the quantum state for the optimum measurement.  The classical control algorithm uses the simultaneous perturbation stochastic approximation (SPSA) method to update the setting $\bm w$ for the variational circuits. 
(c)(d) illustration of the performance contrast between the entanglement-assisted SVM and classical SVM in the parameter space. The labels `+' and `-' denotes the different classes of the channels. The blur is due to measurement uncertainties. While the measurement uncertainty renders classical SVM inaccurate, the entanglement-assisted SVM remains intact.
\label{schematic_M_general_SVM}
}
\end{figure}

To demonstrate the power of SLAEN, we utilize it to construct an entanglement-assisted SVM for physical-layer data classification, shown in Fig.~\ref{schematic_M_general_SVM} (b), and make a comparison with the classical SVM scheme, shown in Fig.~\ref{schematic_M_general_SVM} (a). In both regimes, the interrogation of an object is modeled by channel $\Phi^{(n)}$ that imparts $M$ unitary operations, $\hat{U}(\alpha_1^{(n)})$ to $\hat{U}(\alpha_M^{(n)})$, on the probe state represented by $M$ annihilation operators $\hat{a}_1$ to $\hat{a}_M$. At present, we assume the channels are lossless, but will later add in pure loss channels with transmissivity $\eta$ to account for experimental imperfections. Here, we consider displacement unitaries $\hat{U}\left(\alpha^{(n)}_m\right)\equiv \exp\left(-i\alpha^{(n)}_m \hat{p}\right)$, with the understanding that phase shift can be transformed into field displacement by a Mach-Zehnder interferometer (see Appendix~\ref{sec_convert} for details), which thereby greatly broadens the scope that SLAEN applies to. 

In the above scenario, each channel instance is described by a set of displacements, $\bm \alpha^{(n)}=\left(\alpha^{(n)}_1,\cdots, \alpha^{(n)}_M\right)$. The data acquisition process aims to estimate global features of the object embedded in the channel parameter $\bm \alpha^{(n)}$, while the exact value of $\bm \alpha^{(n)}$ is unknown to the SL tasks. To this end, the classical SVM scheme obtains an classical estimate $\tilde{\bm \alpha}^{(n)}$ for all components of $\bm \alpha^{(n)}$ through local measurements and solves the problem by a classical SVM algorithm~\cite{cortes1995support} supplied with the acquired classical data. In this classical SVM, a hyperplane described by $\bm w \in \mathbb{R}^{M},b \in \mathbb{R}$ is chosen, and
\be
\hat{y}^{(n)}={\rm sign}\left(\bm w \cdot \tilde{\bm \alpha}^{(n)}+b\right)
\label{SVM_classical}
\ee
is used to generate a class label.
Here, ${\rm sign}\left(\cdot\right)$ is the sign function, and the hyperplane is optimized by the classical SVM using training data. A close look at the classical SVM scheme makes one notice that Eq.~\ref{SVM_classical} only relies on a global quantity $\bm w \cdot \bm \alpha^{(n)}$ of the channel, thus measuring each component of $\bm \alpha^{(n)}$ and generating a large amount of classical data is redundant and resource inefficient. To resolve this issue, the SLAEN architecture is adopted to perform data classification at a physical layer. 

Our goal is to optimize the entangled probe state and the measurement setting to minimize the error probability in producing a class label for the object. To this end, we introduce a variational circuit, implemented by a unitary operation $\hat{B}^\dag(\bm w)$ comprised of beam splitters and phase shifters~\cite{Weedbrook_2012}, that acts on a resource squeezed vacuum state $\hat{b}_1$ with mean photon number $N_S$ to generate the entangled probe state. A second variational circuit $\hat{B}(\bm w)$ operates on the channel's output entangled probe state $\hat{a}'_n$'s to generate a single mode $\hat{b}'_1$ for detection. A {\em single} homodyne measurement on $\hat{b}'_1$ yields outcome $f_{\bm w}\left(\Phi^{(n)}\right)$. The intuition of such a design is that the expectation value of $f_{\bm w}\left(\Phi^{(n)}\right)$ equals the  weighted sum $\bm w\cdot \bm \alpha^{(n)}$ (see Appendix~\ref{sec_DS} for details), which is exactly what Eq.~\ref{SVM_classical} requires for SVM classification. Moreover, the multipartite entanglement suppresses the measurement uncertainties in analogy to the distributed sensing scheme~\cite{zhuang2018distributed}. As such, a {\em single} measurement on the detection mode suffices. In contrast, in the classical SVM scheme, one needs to employ $M$ single-mode squeezed vacuum sources and $M$ homodyne measurements to obtain an estimation of the entire vector $\bm \alpha^{(n)}$. As such, SLAEN significantly reduces the resource overhead while boosting the performance.

{\em Training stage.---} Without loss of generality, we first consider a binary classification problem. For a specific variational circuit setting $\bm w$, one generates a class label for the object as $\hat{y}^{(n)}={\rm sign}\left(f_{\bm w}\left(\Phi^{(n)}\right)+b\right)$, where $b$ is tuned in classical post-processing. To train the variational circuits, $N$ quantum channels $\Phi^{(n)}$, each labeled as $y^{(n)}$, are provided. The objective is to find the optimum SVM parameters $\bm w^\star, b^\star$ to minimize the cost function defined by
\be
\calE_\lambda\left(\bm w,b\right)= \sum_{n=1}^N |1-y^{(n)}\left(f_{\bm w}\left(\Phi^{(n)}\right)+b\right)|_++\lambda \|\bm w \|^2.
\label{cost_function}
\ee
Here, $|x|_+$ equals $x$ when $x\ge 0$ and zero otherwise, and $\|\cdot\|$ is a usual two-norm. The term $\lambda \|\bm w \|^2$ avoids over fitting. In this cost function, only the support vectors (SVs) --- points close to the hyperplane described by $\bm w$ and $b$ with $y^{(n)} \left(\bm w\cdot \bm \alpha^{(n)}+b\right)\le 1$---non-trivially contribute to the cost function. The rationale behind the form of the cost function is that in a classification scenario errors mainly occurs on the SVs, thus introducing the deviation of all data points into the cost function is not ideal. 

A typical method to minimize the cost function involves the use of the stochastic gradient descent algorithm~\cite{rosasco2015machine,Rosasco_2015}. However, for a general quantum circuit, an analytical form of the gradient is difficult to obtain due to the complexity associated with the measurement outcome $f_{\bm w}\left(\Phi^{(n)}\right)$. Moreover, direct numerical estimate of the gradient requires a number of measurement-and-prepare steps proportional to the length of $\bm w$. To avoid the inefficiency, we use the simultaneous perturbation stochastic approximation (SPSA) method~\cite{spall2000adaptive,spall1998implementation,SPSA_link}, which only needs to evaluate two points to simultaneously update the estimation of the function minimum and its gradient. 
{Also note that when the measurement term $f_{\bm w}\left(\Phi^{(n)}\right)$ has no noise, it is known that the cost function is convex in $\bm w,b$~\cite{rosasco2015machine,Rosasco_2015} and thus the optimization is efficient with guaranteed convergence. We thus expect that the optimization will generally converge as the number of measurement increases, when noise is not too large.}
With an updated $\bm w$, the transmissivities and phase shifts of the beam splitters in the two variational circuits are adjusted prior to the next training round, until the last training round yields $\bm w^\star$ and $b^\star$.

{\em Utilization stage.---} Having obtained the optimum variational circuits parameterized by $\bm w^\star,b^\star$ from training based on channels $\Phi^{(n)}$'s and their associated class labels $y^{(n)}$'s, we can now perform entanglement-assisted classification on a new measurement outcome $f_{\bm w^\star} (\Phi^{(N+1)})$ from an unknown quantum channel $\Phi^{(N+1)}$. To do so, we use the hyperplane described by $\bm w^\star,b^\star$ to make a decision
\begin{equation}
\label{eq:classificationDecision}
    \hat{y}^{(N+1)}={\rm sign}\left(f_{\bm w^\star} (\Phi^{(N+1)})+b^\star\right).
\end{equation}
By virtue of the reduced measurement error with the multipartite entanglement, a single measurement is sufficient for precise classification.

\begin{figure}
\centering
\subfigure{
\includegraphics[width=0.5\textwidth]{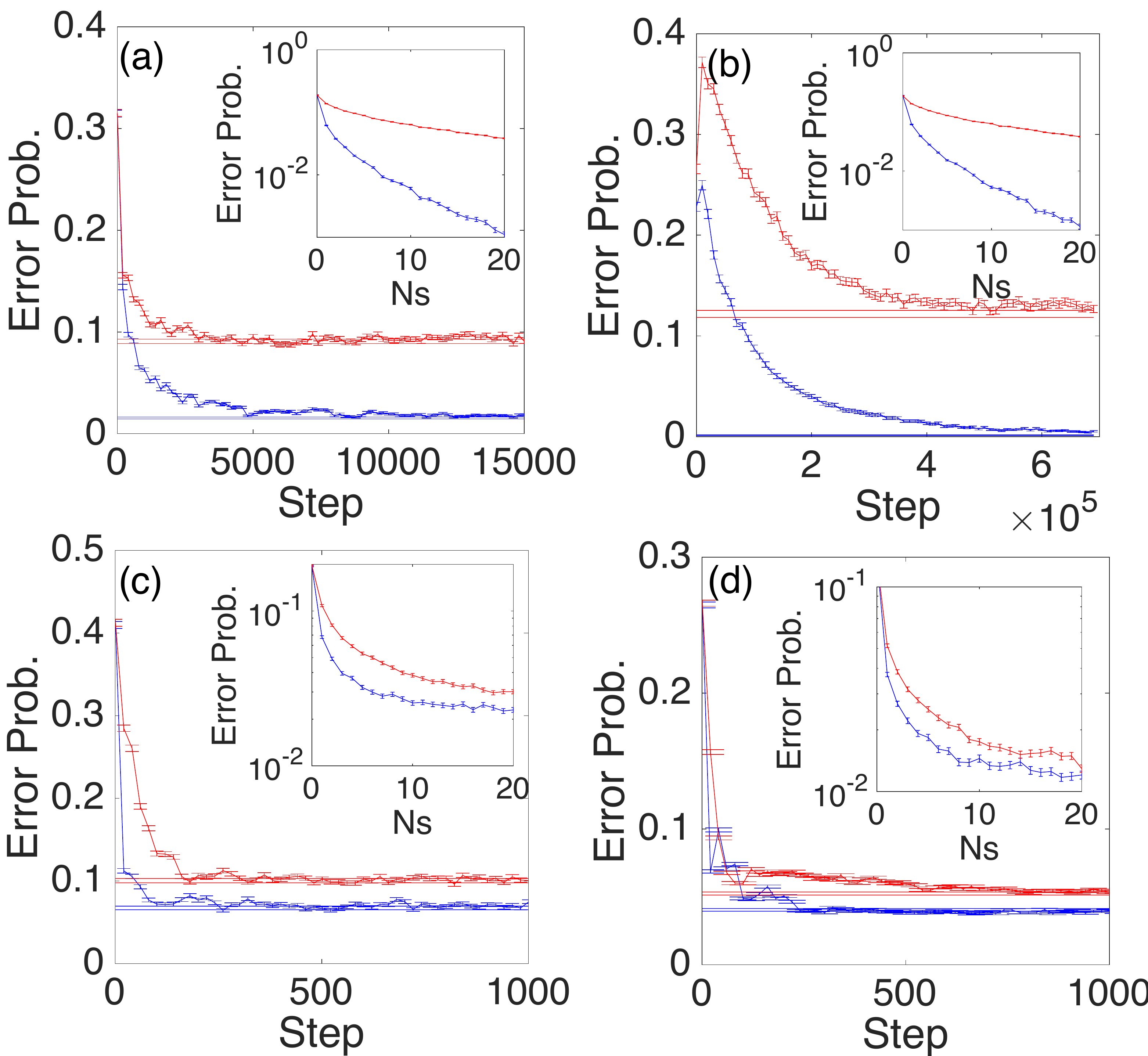}
}
\caption{
Training process and performance for SVM. Data points with error bars: performance of the SVM; Horizontal straight lines: the performance with the perfect hyperplane $\bm w_t$ given, with $\pm$ one standard deviation; Red plots: classical SVM; blue plots: entanglement-assisted SVM. Insets: the scaling of the error probability of the optimum hyperplane with respect to the mean photon number $N_S$. 
(a) $\eta=1$, $N_S=5, \alpha_0=2, M=10$. Number of data points $N=10^3$, $\epsilon=0.1$.
(b) $\eta=1$, $N_S=20, \alpha_0=2, M=100$. $N=10^4$, $\epsilon=0.1$.
(c) $\eta=0.9$, $N_S=1, \alpha_0=1.5, M=3$. $N=500$. $\epsilon=0.2$.
(d) $\eta=0.9$, $N_S=1, \alpha_0=2, M=2$. $N=500$. $\epsilon=0.2$.
In each training step, the channel is randomly chosen from $N$ data points and thus the total training steps can be larger than $N$.
\label{training_M}
}
\end{figure}

{\em SVM simulations.---} To compare the performance of the entanglement-assisted SVM and classical SVM, we simulate the training process with $N$ data points $\left(\bm \alpha^{(n)}, y^{(n)}\right), 1\le n \le N$. Each component $\alpha^{(n)}_m, 1\le m \le M$ of vector $\bm \alpha^{(n)}$ is randomly produced following a uniform distribution in $[-\alpha_0/2,\alpha_0/2]$. A random plane $y=\bm w_t\cdot \bm \alpha$ across the origin is subsequently generated, and true labels by $y^{(n)}={\rm sign}\left(\bm w_t\cdot \bm \alpha^{(n)}\right)$ are assigned to each data point. To tune the difficulty of the problem, we exclude points within distance $\epsilon\ge0$ to the hyperplane from the data set. Smaller $\epsilon$ means the SVs are closer to the hyperplane and thus the problem is more difficult. In the simulation, the measurement outcome is generated by Gaussian-distributed random numbers with mean of $\bm w \cdot \bm \alpha^{(n)}$ and variance given by Eqs.~\ref{delta_alpha_E} and \ref{delta_alpha_P_precise} in Appendix~\ref{sec_DS} for a symmetric case. Since the data are randomly generated, equally distributing the mean photon number of the squeezed state to different sensor nodes is optimum on average, thus we do not perform the costly optimization given by Eq.~\ref{delta_alpha_P_precise} for separable-state sensing. The achieved entanglement-assisted enhancement over the classical SVM scheme however holds for the most general case. Fig.~\ref{schematic_M_general_SVM}(c) and (d) provide a conceptual illustration of the performance contrast between entanglement-assisted SVM and classical SVM (see caption for details). 

To choose a proper regularization parameter $\lambda$ to avoid over-fitting, we generate a test data set in the same way as the training data set. Since the performance is near identical for both data sets, we only show the training data result in the main plots of Fig.~\ref{training_M}. The error probabilities of the SVM at different steps of the training process are plotted in connected points. To ensure convergence of the training process, we evaluate the performance of a perfect classifier with assigned surface $y=\bm w_t\cdot \bm \alpha$ in the horizontal solid lines. Due to the measurement uncertainty, even the perfect classifier suffers from a non-zero error probability. Both the entanglement-assisted (blue) and classical (red) SVMs converge, but the entanglement-assisted SVM enjoys an appreciable error-probability advantage.

We now address the error-probability scaling with respect to the resources, i.e., the number of sensors and the mean photon number of the resource squeezed state, employed in entanglement-assisted SVM. The ultimate error probability is determined by the SVs $\epsilon$-close to the hyperplane. For a given data set, we plot the error probability versus the mean photon number in the inset of Fig.~\ref{training_M}. In the lossless cases, i.e., the transmissivity $\eta = 1$, the insets of Fig.~\ref{training_M} (a)(b) plot the error-probability contrast between the entanglement-assisted SVM and classical SVM. The numerically computed error-probability is remnant of the analytically-derived error-probability scaling advantage for binary channel discrimination depicted in Fig.~\ref{Fig1}(b) in Appendix~\ref{sec_DS}. 

In real applications, the minimum distance of the data points to the hyperplane $\epsilon$ scales with the data dimension, i.e., the number of sensors. In the usual case, each sensor measures a displacement of energy $\sim \epsilon_0^2$, thus the minimum distance scales as $\epsilon\simeq \sqrt{M}\epsilon_0$. The analysis in Appendix~\ref{sec_DS}, which agrees well with the presented numerical results, shows that for large $N_S$ and $M$ the error probability $P_E$ for the entanglement-assisted scheme with homodyne detection scales as $P_E\sim \exp\left(-2N_S M \epsilon_0^2\right)$, whereas the classical SVM scheme has an error-probability scaling of $P_S\sim \exp\left(-4N_S \epsilon_0^2\right)$. This translates to an exponential error-probability advantage with respect to the data dimension, i.e., $P_E\sim P_S^{M/2}$. Such an enormous {performance enhancement} over the classical SVM scheme stems from the utilization of multipartite entanglement to directly extract a global feature of the object in a single-shot measurement, in contrast to the classical SVM scheme's need for individual measurements on all local features. {Note that in the entanglement-assisted SVM and the classical SVM schemes, the number of parameters being optimized in $\bm w,b$ are the same. As such, the enhancement is not due to an increase in the number of fitting parameters. }

In Fig.~\ref{training_M} (c)(d), we consider an entanglement-assisted SVM with less mean photon number ($N_S=1$) and number of sensors ($M=2,3$). We also include a pure loss channel with transmissivity $\eta=0.9$ after each displacement unitary. These parameters represent an entanglement-assisted SVM readily implementable in a bulk-optics platform. Remarkably, despite the saturation of the scaling advantage due to the presence of loss, as shown in the inset plots of Fig.~\ref{training_M} (c)(d), the entanglement-assisted SVM's error-probability advantage survives the loss, demonstrating its tolerance to experimental imperfections.

\section{Entanglement-assisted physical-layer principal-component analyzer}
\label{sec_EPCA}
\begin{figure}
\centering
\includegraphics[width=0.5\textwidth]{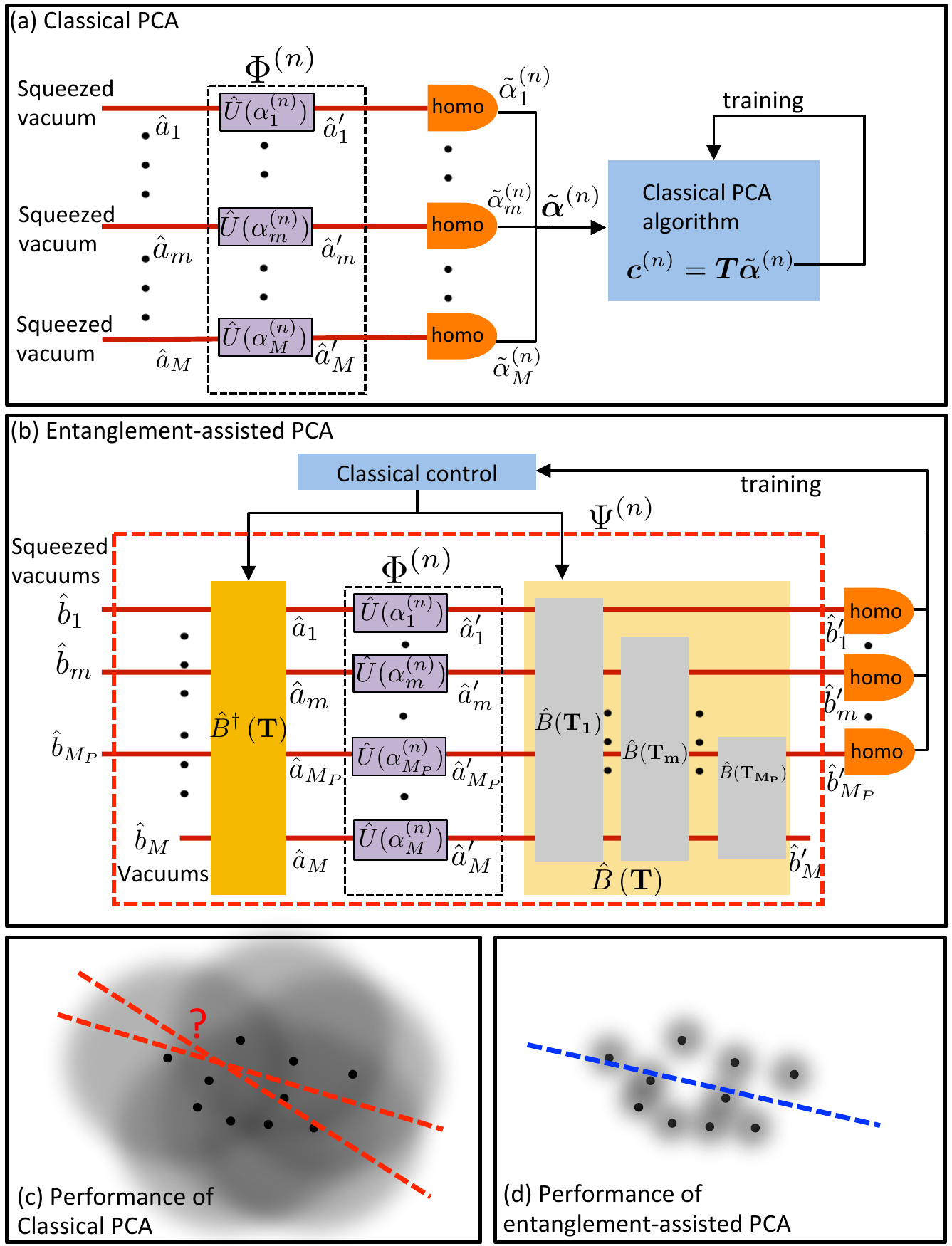}
\caption{(a) schematic of the classical PCA scheme. Separable probes are used to obtain a full noisy classical description $\tilde{\bm \alpha}^{(n)}$ of the unknown channel $\Phi^{(n)}$. Then a classical PCA algorithm finishes the compression task. (b) schematic of the entanglement-assisted PCA. The original channel $\Phi^{(n)}$ with dimension $M$ is reduced to a new channel $\Psi^{(n)}$ with dimension $M_P$. The dimension reduction is carried out by two variational circuits, $\hat{B}(\bm T)^\dag$ and $\hat{B}(\bm T)$ training by a classical algorithm. The training process requires $M_P$ squeezed vacuum states at the input modes $\hat{b}_1$ to $\hat{b}_{M_P}$ and $M_P$ homodyne measurements at the output modes $\hat{b}'_1$ to $\hat{b}'_{M_P}$.
(c)(d) illustration of the performance contrast between the entanglement-assisted PCA and classical PCA in the parameter space. The blur is due to measurement uncertainties. While the measurement uncertainties prevent the classical PCA from finding the principal axis, the entanglement-assisted PCA precisely identifies the principal axis.
\label{schematic_M_general_PCA}
}
\end{figure}
Apart from entanglement-assisted physical-layer SVM, SLAEN can also be leveraged to construct an entanglement-assisted physical-layer PCA as a powerful tool for dimensional reduction and quantum data compression. Utilized in SVM, for example, PCA is particularly useful for data dimension, i.e., the number of input and output modes of $\Phi^{(n)}$, reduction, so that the training of the optimum hyperplane for high-dimensional channels becomes viable (see Fig.~\ref{schematic_M_general_SVM_PCA_combine}). In a classical PCA scheme shown in Fig.~\ref{schematic_M_general_PCA} (a), one needs to input separable probes to measure all data components and obtain a noisy estimation $\tilde{\bm\alpha}^{(n)}\in \mathbb{R}^M$ of the unknown classical description $\bm \alpha^{(n)}\in \mathbb{R}^M$ of $\Phi^{(n)}$. Given the estimated classical description $\tilde{\bm \alpha}^{(n)}\in \mathbb{R}^M$, a classical PCA algorithm identifies the principal components (PCs) stored in a vector $\bm c^{(n)}\in \mathbb{R}^{M_P}$ with dimension $M_P\ll M$ through a linear transform $\bm c^{(n)} =\bm T \tilde{\bm\alpha}^{(n)} $, where $\bm T\in \mathbb{R}^{M_P\times M}$. These PCs are independent and form the large variance subspace of the original data set.

To avoid the inefficient channel tomography, we train a variational circuit to directly produce a reduced-dimension channel $\Psi^{(n)}$ with $M_P$ input modes and $M_P$ output modes. $\Psi^{(n)}$ is described by the principal components $\bm c^{(n)}$. $\Psi^{(n)}$ is obtained by conjugating the original $M$-mode channel $\Phi^{(n)}$ with a beam-splitter array $\hat{B}(\bm T)$, as illustrated in the red box of Fig.~\ref{schematic_M_general_PCA} (b). The construction of the variational circuit is similar to that for entanglement-assisted SVM shown in Fig.~\ref{schematic_M_general_SVM} (b), except that the beam-splitter array for PCA requires the control of weights on $M_P$ modes. The mode reduction is achieved by leaving the $M-M_P$ modes, $\hat{b}_{M_P+1}$ to $\hat{b}_{M}$, in vacuum at the input and discarding $M-M_P$ modes, $\hat{b}'_{M_P+1}$ to $\hat{b}'_{M}$, at the output. Due to the conjugation relation of the two beam splitters, the new channel $\Psi^{(n)}$ remains unitary in the absence of extra loss.

{\em Training stage.---} To train the variational circuits, squeezed vacuum states are injected at the $M_P$ input modes of the channel $\Psi^{(n)}$. A homodyne measurement is performed on each of the $M_P$ output modes, yielding joint measurement outcomes $f_{\bm T}\left(\Phi^{(n)}\right)\in \mathbb{R}^{M_P}$. Since precise knowledge of $\Phi^{(n)}$ is unavailable, we use the gradient-free SPSA method, in lieu of the gradient descent approach~\cite{shamir2016convergence}, to maximize $\|f_{\bm T}\left(\Phi^{(n)}\right)\|^2$ by tuning the beam-splitter array parameter $\bm T$ based on the measurement outcome. {Note that when there is no noise in the measurement, the above cost function is convex with guaranteed convergence.} As shown in Fig.~\ref{schematic_M_general_PCA}, the training process can be performed in a sequential order: the variational circuit is decomposed into $\hat{B}\left(\bm T\right)=\hat{B}\left(\bm T_1\right) \circ \hat{B}\left(\bm T_2\right) \circ \cdots \circ \hat{B}\left(\bm T_m\right) \circ\cdots \circ\hat{B}\left(\bm T_{M_P}\right)$. In doing so, each transform $\hat{B}\left(\bm T_m\right)$ only takes the $m$th to $M$th modes as its input. In the $m$th step, one maximizes the measurement variance of mode $\hat{b}^\prime_m$ by solely tuning $\hat{B}\left(\bm T_m\right)$, with the previous optimized beam-splitter arrays $\hat{B}\left(\bm T_1^\star\right) \circ \cdots \circ \hat{B}\left(\bm T_{m-1}^\star\right)$ unchanged. Executing the process from $m=1$ to $m=M_P$, one obtains the optimum transform $\bm T^\star$ to produce $\Psi^{(n)}$ characterized by the PCs.

{\em Utilization stage.---} The training yields the optimum variational circuits parameterized by $\bm T^\star$. To use the entanglement-assisted PCA, $\hat{B}^\dagger\left(\bm T^\star\right)$ is applied on the $M$ input modes of the channel $\Phi^{(N+1)}$, and $\hat{B}\left(\bm T^\star\right)$ is applied on its $M$ output modes. This results in a reduced-dimension effective channel with $M_P$ modes at both the input and output. 

The reduced-dimensional channel can in turn be used in other SL tasks. Fig.~\ref{schematic_M_general_SVM_PCA_combine} provides an example: the entanglement-assisted PCA is utilized as the pre-processor of the entanglement-assisted SVM. Given an $M$-mode channel $\Phi^{(n)}$, prior to utilizing the entanglement-assisted SVM, one first reduces its dimension to an $M_P$-mode channel $\Psi^{(n)}$ by the entanglement-assisted PCA, as illustrated by the components in the red dashed rectangle. In doing so, the SVM only needs to process a reduced-dimension channel $\Psi^{(n)}$, thus leading to a significant speedup in training the SVM.
\begin{figure}
\centering
\subfigure{
\includegraphics[width=0.205\textwidth]{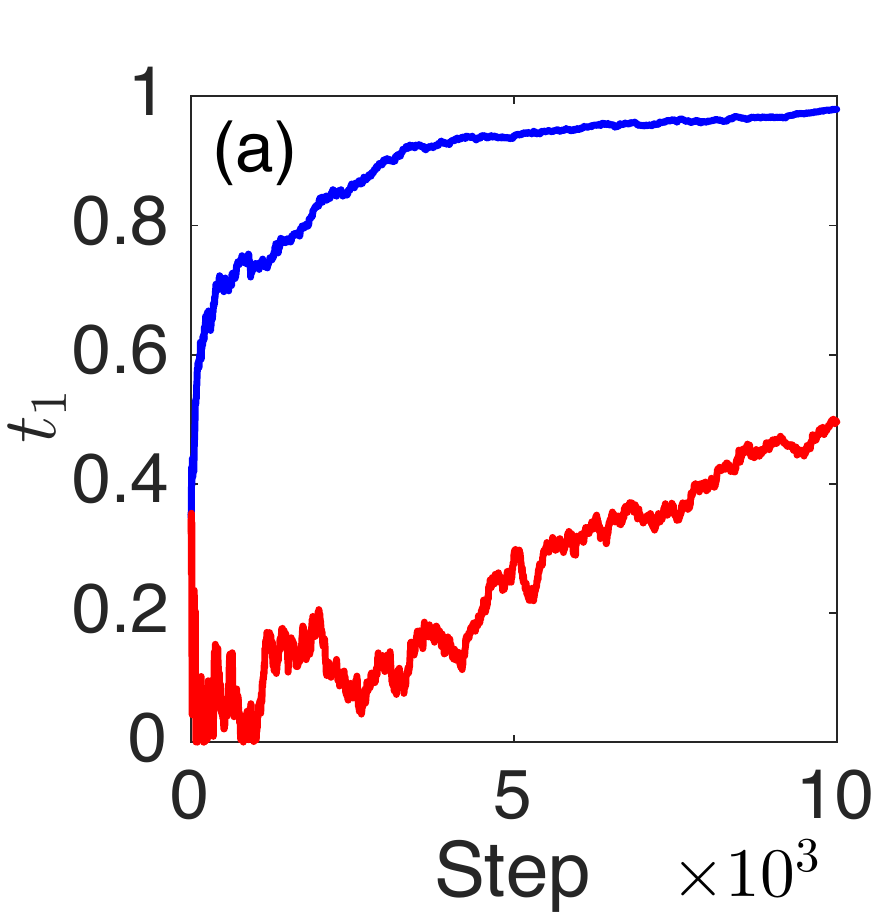}
}
\subfigure{
\includegraphics[width=0.2\textwidth]{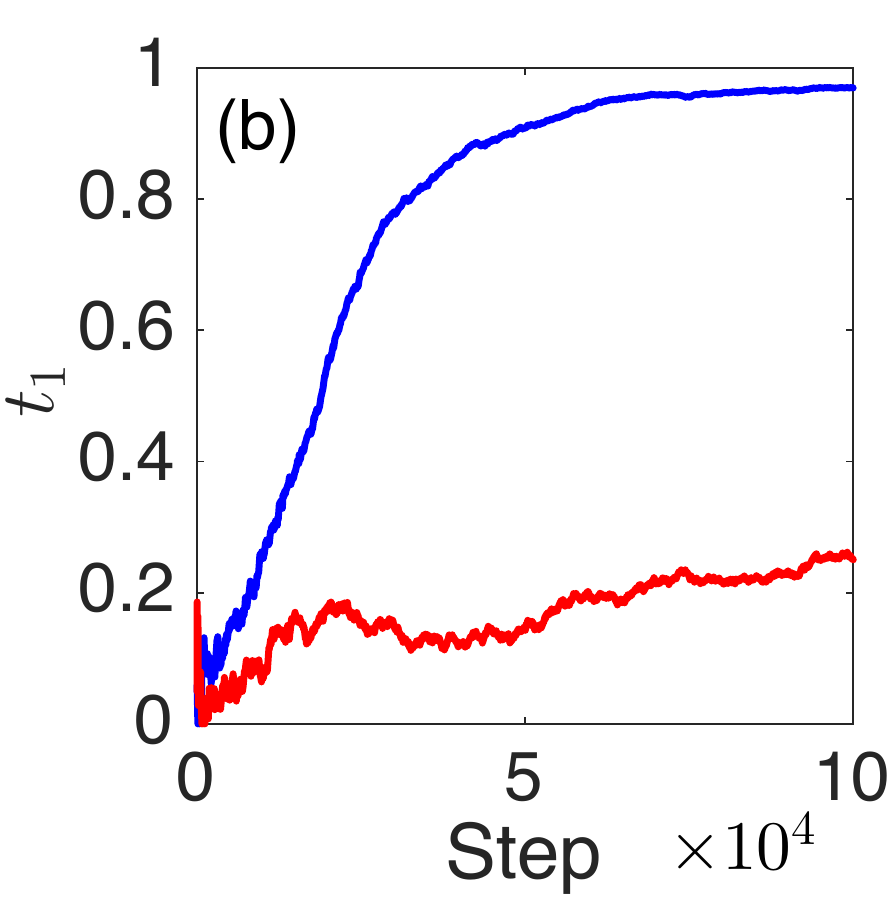}
}
\caption{
Training process of the entanglement-assisted PCA. Blue plots: performance of entanglement-assisted PCA; red plots: the performance of classical PCA. (a) $M=20$, $N_S=1$, $\alpha_0=0.3$, $P=20$. (b) $M=100$, $N_S=1$, $\alpha_0=0.4$, $P=100$.
\label{PCA_performance}
}
\end{figure}

{\em PCA simulations.---} We demonstrate the training of the first variational circuit, the beam-splitter array $\hat{B}\left(\bm T_1\right)$, to obtain the reduced-dimension channel associated with the first PC. The training only needs a single-mode squeezed vacuum state with mean photon number $N_S$ at the input. We randomly generate channels $\Phi^{(n)}$ with zero-mean Gaussian-distributed $\bm \alpha^{(n)}$'s and simulate the quantum noise in the measurement outcome. We set the covariance matrix to $\alpha^2_0{\rm Diag}\left[P,1,1,\cdots,1\right]$ with $P>1$, thus the true PC is obtained by the transform $\bm T_1^\star=[1,0,0,\cdots,0]$. Since we initialize the parameters $\bm T_1$ randomly, a diagonal covariance matrix is general. We plot the first element $t_1$ of the normalized vector $\bm T_1$ during training and compare it with $\bm T_1^\star$. Similar to what has been done in the entanglement-assisted SVM, we compare the performance of the entanglement-assisted PCA (blue) with the optimum classical PCA (red) in Fig.~\ref{PCA_performance}. The result shows that while the classical PCA struggles to find the PC, the entanglement-assisted PCA converges very close to the actual PC. This is because vacuum noise dominates the measurements in classical PCA due to a small variance of $\alpha_0^2$, leading to a close to symmetric variances over all modes. In contrast, the entanglement-assisted PCA leverages the multipartite entanglement to reduce the vacuum noise and thus is able to obtain the correct PC. Fig.~\ref{schematic_M_general_PCA}(c) and (d) conceptually illustrate that entanglement-assisted PCA is able to solve classically intractable problems. 

\begin{figure}
\centering
\includegraphics[width=0.5\textwidth]{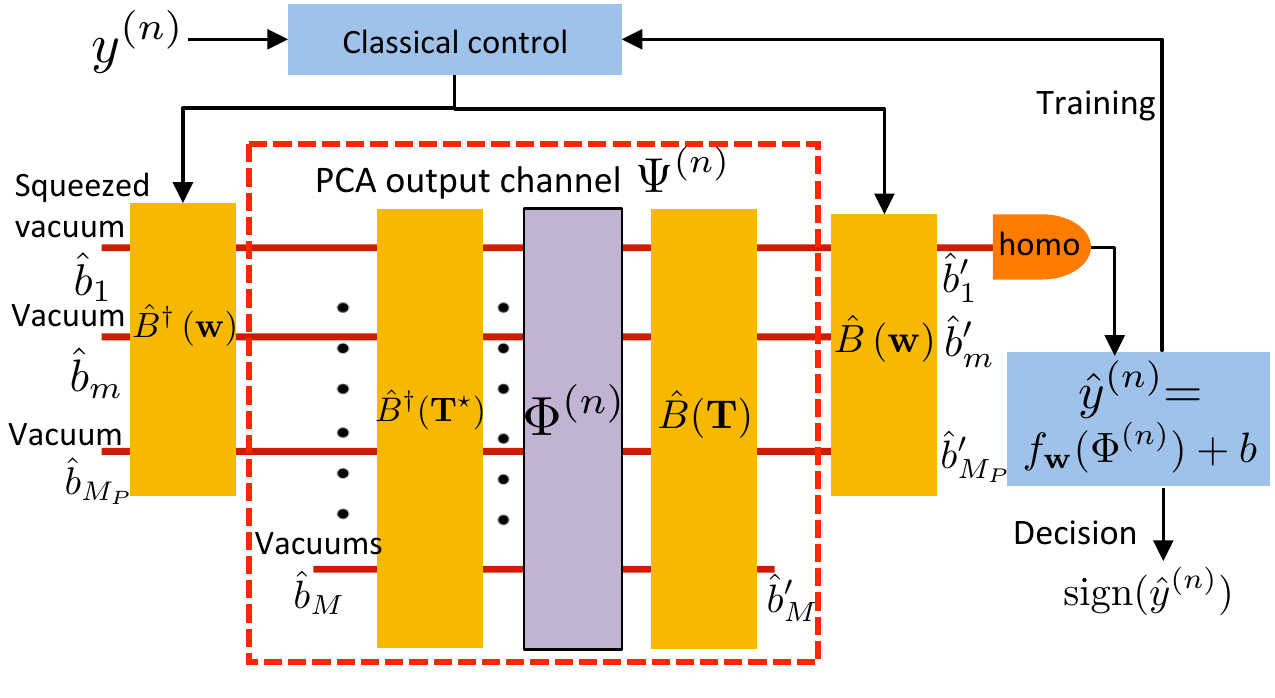}
\caption{
A combined scheme with entanglement-assisted SVM and PCA. Using multipartite entangled probe states, the PCA is training to reduce the original $M$-mode quantum channel $\Phi^{(n)}$ (purple rectangle) to a $M_P$-mode quantum channel $\Psi^{(n)}$ (red dashed box), which is in turn used by the entanglement-assisted SVM to carry out data classification tasks.
\label{schematic_M_general_SVM_PCA_combine}
}
\end{figure}

\section{Discussions}

In this paper, we utilize SLAEN to construct the entanglement-assisted SVM and PCA. SLAEN only requires off-the-shelf components and are robust against noise, thus representing a NISQ paradigm for demonstrating a practical {entanglement enhancement} over classical SL schemes without entanglement. SLAEN measures field-quadrature displacements, but it can be generalized to cope with other displacements on canonical conjugate variables including time and frequency, angular momentum and angle, and number and phase.

Our entanglement-assisted SVM opens new avenues for ultrasensitive measurements in biological, thermal, and mechanical systems. A key ingredient for the entanglement-assisted SVM is transducers that convert the information carried on the samples into phase modulation. For example, electro-optic devices can convert radio-frequency (RF) signals into refractive index shift of optical materials, which in turn impart phase modulation on light~\cite{fan2018superconducting}. In optical biosensors, the evanescent wave of the probe light interacts with the sample, and the induced phase shift serves as a means to identify the density and species of the bio-molecule~\cite{fernandez2016last}. These phase shifts, in turn can be transformed into displacement through a Mach-Zehnder interferometer (see Fig.~\ref{phase_to_disp} in Appendix~\ref{sec_convert}), where for the single-mode case squeezed state is known to be beneficial~\cite{escher2011general,demkowicz2013fundamental}. In such a scenario, the entanglement-assisted SVM enables much more efficient and sensitive classification on data acquired by phase sensing than what sensors based on classical light can afford. 

In the given sensing examples, the transducers can be engineered to introduce nonlinearity when converting the physical parameters being probed into phase modulation on the entangled photons. In RF sensing, the intrinsic nonlinearity of the electro-optic devices can be leveraged. Likewise, by properly designing the waveguide geometry, the phase shift induced by optical biosensors can be engineered to be nonlinearly dependent on the density of the bio-molecules. The introduced measurement nonlinearity would be a key to implement entanglement-assisted general-kernel SVM. 

Before closing, we point out that the proposed entanglement-assisted SVM applies in a very different scenarios than existing proposals for quantum SVMs~\cite{rebentrost2014quantum}. 
By transforming the SVM cost function to a quadratic form~\cite{suykens1999least}, quantum speedup is envisaged with mechanisms similar to a linear system of equations~\cite{HHL}, assuming qRAMs are available. Determining the exact quantum advantage of such schemes, however, is still ongoing research. Interestingly, more recent studies of quantum-inspired classical algorithms have demonstrated capabilities of solving similar problems in certain scenarios~\cite{gilyen2018quantum,tang2018quantum1,tang2018quantum2,arrazola2019}. Moreover, due to the channel noise, such quantum SVM algorithms is not directly applicable to the channel learning problems. 

A few future developments for SLAEN would make it more widely applicable. First, the binary-hypothesis entanglement-assisted SVM can be generalized to a multi-hypothesis SVM by introducing a series of binary classification tasks~\cite{rifkin2004defense,park2007efficient} or by adding multiple hyperplanes~\cite{weston1999support}. The current SLAEN architecture can be adapted to accommodate either case. Second, introducing nonlinear maps into SLAEN would make it capable of dealing with nonlinear data classification and compression problems. Third, an extension of the CV sensing scheme to incorporate DV sensing of phase rotations is another interesting open problem. Ref.~\cite{eldredge2018optimal} presents a proposal for measuring a weighted average through controlling the integration time on each qubit. It is open whether there are ways of implementing the measurement for given fixed unitary phase rotations. While direct measurements of a weighted average appears difficult, classification without resorting to direct intuitions has been considered~\cite{havlicek2018supervised,schuld2018quantum,biamonte2017quantum,dunjko2018machine}. 
Fourth, since quantum-state preparation in qRAMs involves quantum measurements that lead to noisy data, with some modifications, SLAEN may be potentially utilized to enhance the performance of quantum machine-learning schemes based on qRAMs.

\begin{acknowledgements}

We thank helpful discussions with Jeffrey Shapiro, Umesh Vazirani, Zeph Landau, Saikat Guha, and William Clark. Q.Z. is supported in part by the Office of Naval Research Award No.~N00014-19-1-2189 and the GeoFlow Grant No. DE-SC0019380. Z.Z. is supported in part by the Office of Naval Research Award No.~N00014-19-1-2190. Q.Z. and Z.Z. are grateful for the support from the University of Arizona. 

\end{acknowledgements}

\appendix

\section{Entangled sensor network for channel discrimination}
\label{sec_DS}

\begin{figure}
\centering
\subfigure{
\includegraphics[width=0.29\textwidth]{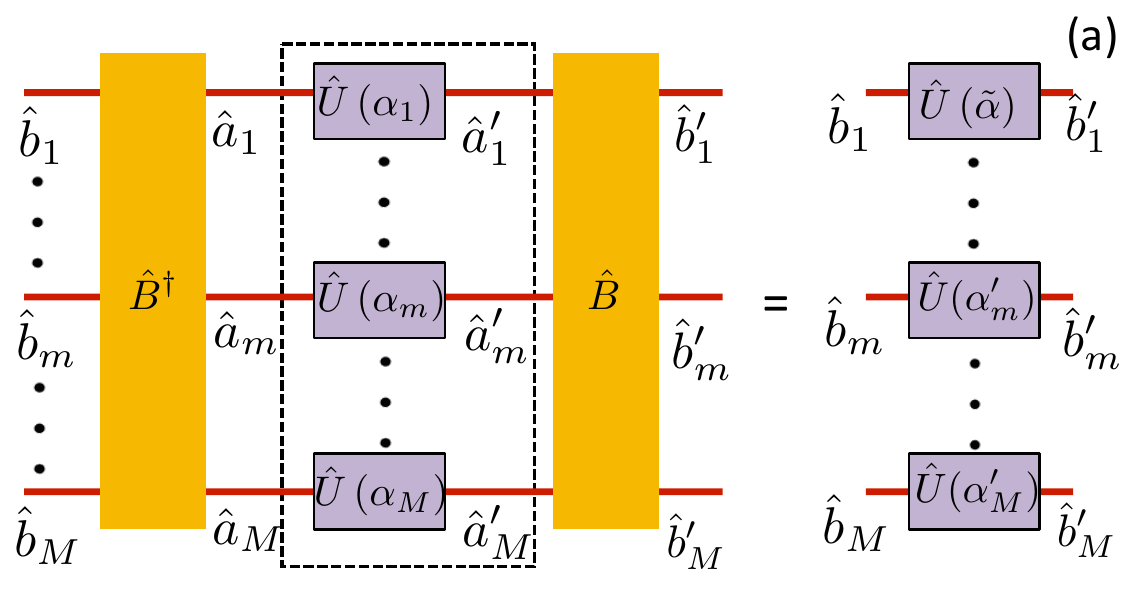}
}
\subfigure{
\includegraphics[width=0.16\textwidth]{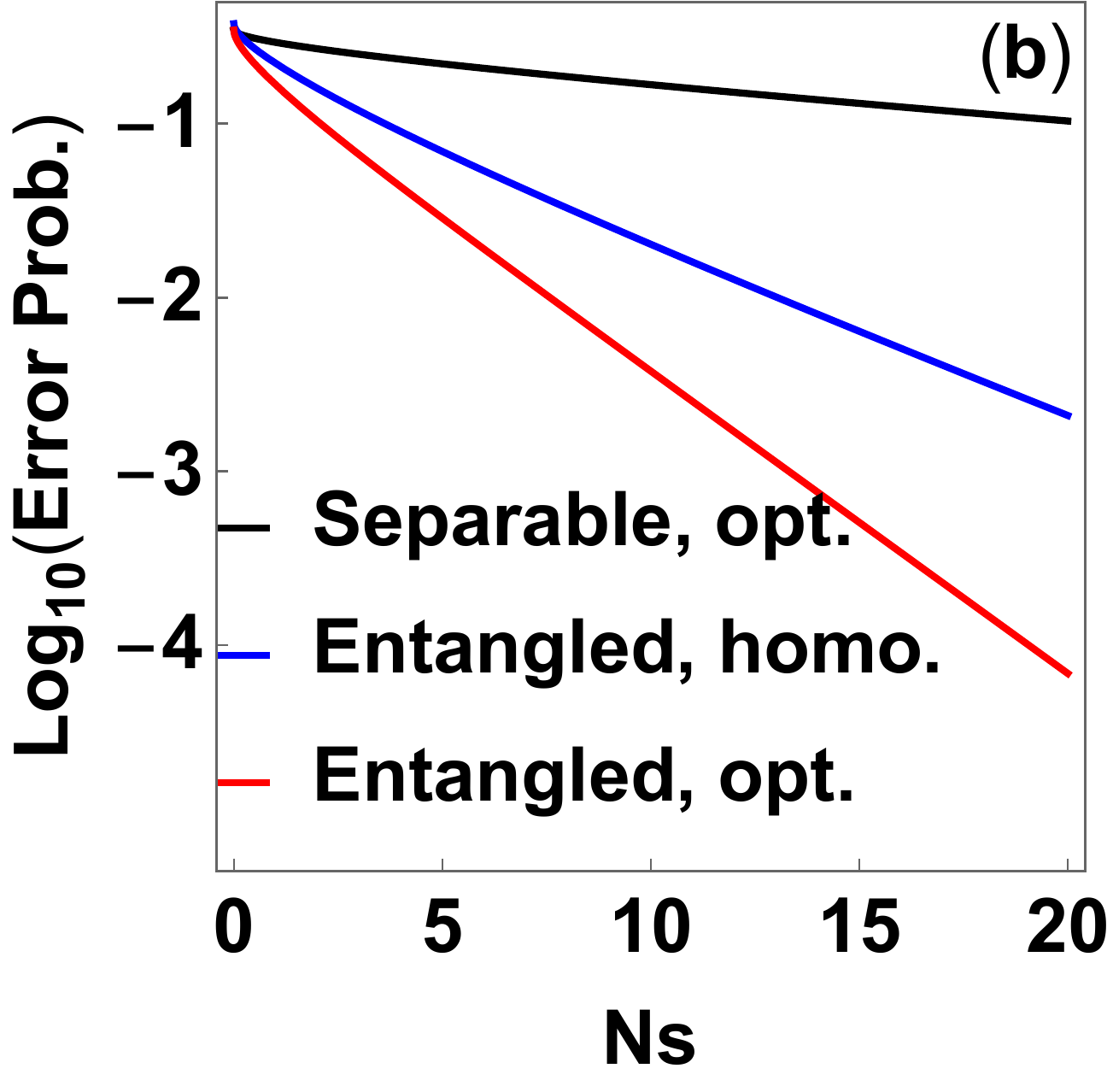}
}
\caption{
(a) Interferometer structure for amplitude mixing in a distributed sensing scheme.
(b) Performance of the symmetric case, $M=10$ and $\alpha_m=0.1, 1\le m \le M$.
\label{Fig1}
}
\end{figure}

A key component in the CV distributed sensing scheme~\cite{zhuang2018distributed} is an interferometer structure formed by two beam-splitter arrays. By conjugating an array of displacement unitaries with a suitable beam-splitter unitary $\hat{B}$ (see Fig.~\ref{Fig1}(a)), we can mix the amplitudes of displacements:
\be 
\hat{B}^\dagger
\left(
\otimes_{m=1}^M
\hat{U}\left(\alpha_m\right)
\right)
\hat{B}
=\hat{U}\left(\tilde{\alpha}\right)\otimes 
\left( 
\otimes_{\ell=2}^M
\hat{U}\left(\alpha_m^\prime\right)
\right),
\ee
where $\tilde{\alpha}=\sum_m w_m \alpha_m$ is the effective displacement between modes $\hat{b}_1$ and $\hat{b}_1^\prime$. The weights $w_m$'s are normalized, i.e., $\sum_{m=1}^M w_m^2=1$. Note that an arbitrary choice of weights $w_m$'s only requires $M$ two-mode beam splitters to realize. Suppose we use a squeezed vacuum state with squeezing parameter $r$ at mode $\hat{b}_1$ as the input, then we can measure $\tilde{\alpha}$ more precisely on mode $\hat{b}_1^\prime$. In this way, the input modes $\hat{a}_1,\cdots, \hat{a}_M$ to the displacement channel are in fact entangled. The original entanglement-assisted distributed sensing scheme measures the weighted average $\bm \omega \cdot \bm \alpha^{(n)}$ in presence of a detection loss $\eta$ to a precision 
\be
\delta \alpha^E_\eta=\left(\eta g\left(N_S\right)+1-\eta\right)^{1/2}/2,
\label{delta_alpha_E}
\ee
where $g\left(n\right)\equiv 1/\left(\sqrt{n+1}+\sqrt{n}\right)^2$ and $N_S$ is the total mean photon number; while a scheme using separable squeezed states for modes $\hat{a}_1,\cdots, \hat{a}_M$ has precision given by a constrained optimization, 
\be
\delta \alpha_{ \eta}^P=\min_{\sum_{m=1}^M N_m=N_S }
\left[\eta\sum_{m=1}^M\left(w_m^2 g(N_m)\right)+1-\eta\right]^{1/2}/2,
\label{delta_alpha_P_precise}
\ee
with the total mean photon number $N_S$ as a constraint. In the lossless case, one can prove that the above entangled scheme is optimum among all schemes and the above separable scheme is also optimum among all schemes without using entanglement~\cite{xia2018repeater}. Moreover, in the lossy case, the optimality of both schemes still holds when we restrict the probe states to be Gaussian~\cite{zhuang2018distributed}.
For the equal weight case, the minimum $\delta \tilde{\alpha}_{ \eta}^P$ is obtained by setting $N_m=N_S/M, 1\le m \le M$ in Eq.~\ref{delta_alpha_P_precise}. As shown in Ref.~\cite{zhuang2018distributed}, when the mean photon number per mode $N_S/M$ is fixed, in the lossless case ($\eta=1$) the entangled scheme has a Heisenberg-scaling precision ($\delta \alpha^E_{\eta}\sim 1/M$), while the classical scheme obeys the standard quantum limit ($\delta \alpha^P_{\eta}\sim 1/\sqrt{M}$). Moreover, despite that the Heisenberg scaling is destroyed by loss ($\eta<1$), the advantage vouchsafed by entanglement survives.

To understand how the above advantage translates to channel-classification problems, we calculate the error probability performance of a simple binary channel discrimination task between displacements $\{\bm 0, \bm \alpha\}$ with given prior probabilities $\{\pi_0,\pi_1\}$ in the lossless ($\eta=1$) case. For the case of $\bm 0$, the transformed displacements are all zero regardless of the beam splitters choice. For the case of $\bm \alpha$, we have $\tilde{\alpha}=\|\bm \alpha\|$ while all other displacements are zero, by choosing the beam-splitter array in Fig.~\ref{Fig1}(a) to have $w_m={\alpha_m}/ \|\bm \alpha\|$.
This can always be ensured as follows. First, one performs a beam-splitter transform between $\hat{b}_1$ and $\hat{b}_2$ to produce the mode $\frac{\alpha_1}{\sqrt{\alpha_1^2+\alpha_2^2}} \hat{b}_1+\frac{\alpha_2}{\sqrt{\alpha_1^2+\alpha_2^2}} \hat{b}_2$. Afterwards, one performs another beam-splitter transform on the previous output mode and $\hat{b}_3, \cdots, \hat{b}_M$. Since beam splitters conserve the mean photon number and the effective displacement on $\hat{b}_1$ is $\|\bm \alpha\|$, all the other effective displacements are zero.

In this way, the muti-mode problem is reduced to a single-mode problem of determining whether a displacement unitary has amplitude zero or $\|\bm \alpha\|$. In Appendix~\ref{App_opt_Gaussian}, we show that the optimum input Gaussian state that minimizes the Helstrom limit~\cite{Helstrom_1976} is a single-mode squeezed vacuum state and the corresponding error probability is
\ba
&P_{\{\pi_0,\pi_1\}}^E=\left[1-\sqrt{1-4\pi_0\pi_1\exp\left(-e^{2r} \|\bm \alpha\|^2\right)}\right]/2
\nonumber
\\
&\simeq \pi_0\pi_1\exp\left(-e^{2r} \|\bm \alpha\|^2\right),
\label{Helstrom}
\ea
where $\sinh^2\left(r\right)=N_S$ is the mean photon number. Thus, in terms of the error probability, the considered entangled scheme is the optimum and the considered classical scheme is the optimum among all {\em Gaussian} schemes without using entanglement. In general, non-Gaussian inputs may lead to better performances, but the advantage of an entanglement-assisted sensing scheme still holds.

Having obtained the optimum Gaussian input state that minimizes the Helstrom limit, now we derive the corresponding measurement scheme to achieve the Helstrom limit.
The construction is as follows. First, one performs a squeezing of amplitude $-r$ on $\hat{b}_1^\prime$ to obtain the coherent state $\ket{e^r \|\bm \alpha\|}$ or $\ket{0}$; Next, one applies a slicing Dolinar receiver~\cite{dolinar_processing_1973} scheme on the output. The optimality of the above scheme simply follows from the optimality of the Dolinar receiver.
However, this construction requires feed-forward and photon number counting. A more practical scheme is to perform a homodyne detection with a maximum-likelihood decision rule, where the error probability can be evaluated by the Gaussian error function. For the $\pi_1=\pi_0=1/2$ case, the formula is simple,
\be 
P_{\{1/2,1/2\}}^{E,Het}=\frac{1}{2}{\rm Erfc}\left[\frac{\|\bm \alpha\|}{\sqrt{2}e^{-r}}\right] \sim \exp\left[-e^{2r}\|\bm \alpha\|^2/2\right].
\ee
Comparing to Eq.~\ref{Helstrom}, it is only a factor of two worse in the error exponent in the asymptotic limit.

Now we compare the performance of the entangled scheme with the optimum Gaussian separable-state scheme. To ensure a fair comparison, we set the total mean photon number to be $N_S=\sum_{m=1}^M N_m$, where $N_m\equiv \sinh^2\left(r_m\right)$ is the mean photon number of the input to the $m$th displacement. Given the constraint on the mean photon number $N_m$, for each mode a single-mode squeezed state is the optimum Gaussian input state, thus to minimize the Helstrom limit, one needs to optimize the distribution of mean photon number to minimize the overlap
$
|\braket{\psi_0^M|\psi_1^M}|^2= \exp\left(-\sum_{m=1} ^Me^{2r_m}\alpha_m^2\right),
$
where $\psi_k^M, k=0,1$ are the $M$-mode joint output states in each hypothesis.
Given the minimum $E_{\rm min}$ of $|\braket{\psi_0^M|\psi_1^M}|^2$, the Helstrom limit of the error probability is $P^{S}_{\pi_0,\pi_1}=\frac{1}{2}\left[1-\sqrt{1-4\pi_0\pi_1 E_{\rm min}}\right]$. In general, there is no closed form solution. Because $ e^{2r_m}$, as a function of $N_m$, is concave, so for equal $\alpha_m$'s it is optimum to evenly distribute the photons. Moreover, for large $N_S$'s concentrating most photon number on the mode with the largest $\alpha_m$ is close to the optimum.

From the above analysis, we see the entanglement-assisted scheme is strictly better than the classical scheme. the advantage is the most significant under $\sum_{m=1}^M\alpha_m^2\gg \max_{1\le m \le M} \alpha_m^2$. Asymptotically, for a fixed $M$, a large $N_S$, and $\bm \alpha=\left(\alpha,\cdots,\alpha\right)$, the error exponents for the classical scheme, the entanglement-assisted scheme with homodyne measurements, and the entanglement-assisted scheme with the optimum measurements are $4N_S \alpha^2$, $2N_S M \alpha^2$, and $4 N_S M \alpha^2$. This shows a substantial advantage for large $M$'s. We give a numerical example in Fig.~\ref{Fig1}(b) for a moderate size $M=10$ and show that the advantage is already appreciable.

\section{Optimum Gaussian state for single-mode displacement discrimination}
\label{App_opt_Gaussian}

Now we consider the single-mode channel discrimination problem of two displacement operators $\hat{U}\left(0\right)$ and $\hat{U}\left(\alpha\right)$. We show that the input Gaussian state that minimizes the Helstrom limit of the error probability is the single-mode squeezed vacuum state.

\begin{figure}
\centering
\includegraphics[width=0.3\textwidth]{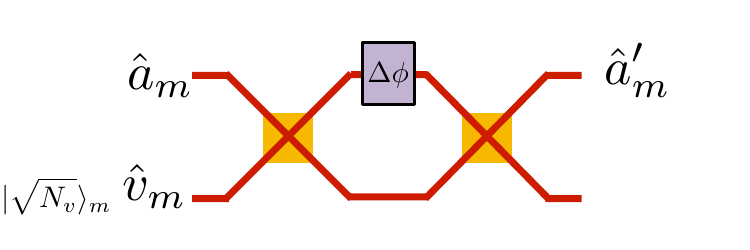}
\caption{
Schematic of conversion between phase shift and displacement through a Mach-Zehnder interferometer.
\label{phase_to_disp}
}
\end{figure}

We first consider an input mode entangled with an ancilla mode. Due to the ancilla, it suffices to consider an input-ancilla joint pure state $\ket{\psi}$. Then in each hypothesis we have the output state $\ket{\psi_0}=\ket{\psi}$ and $\ket{\psi_1}= \hat{U}\left(\alpha\right)\otimes \hat{I}\ket{\psi}$. 
Given the prior probabilities $\pi_0,\pi_1=1-\pi_0$,
the minimum error probability in the discrimination of the two outputs is given by the Helstrom limit
\be
P_{\{\pi_0,\pi_1\}}\left(\psi_0,\psi_1\right)=\frac{1}{2}\left[1-\sqrt{1-4\pi_0\pi_1|\braket{\psi_0|\psi_1}|^2}\right].
\label{helstrom_pure}
\ee
Thus we aim to minimize $|\braket{\psi_0|\psi_1}|^2=|\braket{\psi|\left(\hat{U}\left(\alpha\right)\otimes \hat{I}\right)|\psi}|^2$, subject to certain constraints on the input state $\ket{\psi}$.

While the optimization is in general difficult, we can optimize over Gaussian states to obtain the optimum Gaussian input. For pure two-mode Gaussian states with the covariance matrix $V$ and mean $\mu_0=(0,0,0,0)$ and $\mu_1=(2\alpha,0,0,0)$, the overlap is~\cite{Banchi_2015,marian2016quantum}
\be
|\braket{\psi_0|\psi_1}|^2=\exp\left[-\frac{1}{4}(\mu_1-\mu_0)^T V^{-1} (\mu_1-\mu_0)\right].
\ee 
The above equation indicates that minimizing the state overlap is equivalent to minimizing the variance of the first quadrature, as expected from intuition.

The resource in our task is the mean photon number of the first mode. Given the uncertainty principle $\delta^2 q\cdot \delta^2 p\ge 1$ and energy constraint $\delta^2 q+ \delta^2 p\le N_S$, we conclude that it is optimum to use a single-mode squeezed vacuum, without requiring entanglement assistance. Thus, the optimum general Gaussian state for single-mode displacement discrimination is a single-mode squeezed vacuum state.

The corresponding error probability can be obtained, based on the covariance matrix
$V={\rm Diag}\left[e^{-2r},e^{2r},1,1\right]$ and the total mean photon number constraint
$
N_S=\sinh^2\left(r\right)$, as
\ba
P_{\{\pi_0,\pi_1\}}&=&\frac{1}{2}\left[1-\sqrt{1-4\pi_0\pi_1\exp\left(-e^{2r} {\alpha}^2\right)}\right]
\nonumber
\\
&\simeq &\pi_0\pi_1\exp\left(-e^{2r} {\alpha}^2\right).
\ea

\section{Conversion between displacement sensing and phase sensing}
\label{sec_convert}
It suffices to illustrate the displacement-phase conversion using a single-mode case in Fig.~\ref{phase_to_disp}. Consider a phase shift $\Delta \phi\ll1$. When sandwiched between two beam splitters, with one of the input mode $\hat{v}_m$ in a coherent state $\ket{\sqrt{N_v}}_m$, the phase shift leads to a mode transform (to the first order) as
\be 
\hat{a}_m^\prime=(1-i\Delta \phi/2)\hat{a}_m+i\hat{v}_m\Delta \phi/2,
\ee 
where $\Delta\phi$ is embedded in a field-quadrature displacement $\alpha = i\sqrt{N_v}\,\Delta\phi/2$ of $\hat{a}'_m$. To generalize, when a sensor array measures multiple spatiotemporal phase shifts, the output signals will be in a form of multimode displacement.

%

\end{document}